\documentclass[11pt,twoside]{article}
\usepackage{asp2004}
\usepackage{psfig}
\usepackage{epsf}
\usepackage{graphics}
\usepackage{lscape}
\def\edcomment#1{\iffalse\marginpar{\raggedright\sl#1\/}\else\relax\fi}
\reversemarginpar

\def\arcs{\hbox{$^{\prime\prime}$}}

\begin{document}
\title{Magnetic fields of B and Herbig Ae stars
measured with FORS1 at the VLT}
\author{S. Hubrig, T. Szeifert}
\affil{European Southern Observatory, Casilla 19001, Santiago 19, Chile}
\author{P. North}
\affil{Laboratoire d'Astrophysique de l'Ecole Polytechnique F\'ed\'erale
de Lausanne, Observatoire,
CH-1290~Chavannes-des-Bois, Switzerland}
\author{M. Sch\"oller}
\affil{European Southern Observatory, Casilla 19001, Santiago 19, Chile}
\author{R.V. Yudin}
\affil{Central Astronomical Observatory of the Russian Academy of Sciences at Pulkovo, 196140 Saint-Petersburg, Russia}

\begin{abstract}
We present the results of a magnetic survey with FORS\,1 in polarimetric mode
of a sample of B and Herbig Ae stars with previously undetected magnetic
fields.
For the first time a mean longitudinal magnetic field at a level higher
than 3\,$\sigma$ has been detected in one normal B star, two HgMn stars,
one PGa star and four SPB stars.
The observations of the three Herbig Ae stars (also known as Vega-like stars) reveal a definite
longitudinal  magnetic field
in the star HD\,139614 at 4.8\,$\sigma$ level:
$\left<{\cal B}_z\right>$=$-$$450\pm93$\,G.
This is the largest
magnetic field ever diagnosed for a Herbig Ae star.
A hint of a magnetic field is found in the other two stars,
HD\,144432 and HD\,144668, for which
magnetic fields are measured at the $\sim$1.6\,$\sigma$ and $\sim$2.5\,$\sigma$ 
level respectively.
\end{abstract}
\thispagestyle{plain}


\section{Observations}

The multi-mode instrument FORS\,1 is equipped with
polarization analyzing optics comprising super-achromatic half-wave and
quarter-wave phase retarder plates and a Wollaston prism with a beam
divergence of 22\arcs{} in standard resolution mode.
We have recorded circular polarization spectra
using GRISM 600B and an 0.4\arcs{} slit.
The observed spectral range ($\simeq$ 345-590\,nm) includes all Balmer
lines from H$_\beta$ to the Balmer jump. 
The major advantage of using low-resolution spectropolarimetry 
with FORS\,1 is that polarization can be detected in relatively fast rotators as we
measure the field in the hydrogen Balmer lines.

Our sample of B stars consisted of various groups including normal B stars,
HgMn stars, He weak Si stars, PGa stars and Slowly Pulsating B (SPB) stars.
Their position in the H-R diagram obtained from accurate Hipparcos
parallaxes ($\sigma(\pi)/\pi<0.2$) and
using photometric data in the Geneva and Str\"omgren systems for the
determination of the effective temperatures is shown in Fig.\,1 (left).
The basic data together with the measured magnetic fields are presented
in Table\,1.

Our sample included also three Herbig Ae stars which sometimes 
in the literature are called as Vega-like stars. As potential progenitors of 
the magnetic Ap stars, Herbig Ae stars provide an excellent
opportunity to study the early evolution of magnetic fields in stars of same 
masses. A detection of magnetic fields in these stars is especially important in
view of our recent results that Ap magnetic stars of mass
below 3\,$M_\odot$ are significantly evolved
and concentrated towards the centre of the main-sequence band, and
practically no magnetic star of mass below 3\,$M_\odot$ 
can be found close to the zero-age main sequence (ZAMS)
(\citet{Hu00}, \citet{Hu04b}). The search for magnetic fields and the study of 
their structure in the pre-main sequence
counterparts is a crucial step towards understanding of the origin of the
magnetic fields in stars of intermediate mass.
The observations of three Herbig Ae stars, HD\,139614, HD\,144432 and
HD\,144668, have been carried out with FORS\,1 in September 2003.
Spectra of these stars in integral light in the spectral region around the Ca\,II K
line and close-by H Balmer lines
are presented in Fig.\,1 (right). The corresponding Stokes V spectra are shown in
Fig.\,2 (left). The last two bottom panels in these figures show the spectra of the
non-magnetic HgMn star HD\,175640
observed during the same night as the Herbig Ae stars and 
of the classical Ap star HD\,94660 with a well-defined longitudinal field
of $-2$\,kG, occasionally observed by us to check a proper instrument 
functionality.
The assessment of the longitudinal magnetic field using FORS\,1 spectra
is achieved by measuring the
circular polarization of opposite sign induced in the wings of broad lines,
such as Balmer lines, by the Zeeman effect.
Using Balmer lines from H$_\beta$ to H$_{16}$ and Ca\,II K for
the measurements, we derived for the star HD\,139614 the mean longitudinal
field $\left<{\cal B}_z\right>$=$-$$475\pm94$~G. For the stars HD\,144432
and HD\,144668 we measured respectively
$\left<{\cal B}_z\right>$=$-$$94\pm60$\,G and
$\left<{\cal B}_z\right>$=$-$$118\pm48$\,G.

\begin{table}
\caption{Stellar parameters of all stars in our sample.}
{\scriptsize
\begin{tabular}{rrcccrrrcl}
\multicolumn{1}{c}{HD} &
\multicolumn{1}{c}{mass} &
\multicolumn{1}{c}{log $T_{\rm eff}$} &
\multicolumn{1}{c}{log $L$} &
\multicolumn{1}{c}{log $g$} &
\multicolumn{1}{c}{$R/R_{\odot}$} &
\multicolumn{1}{c}{$d$ [pc]} &
\multicolumn{1}{c}{$B_{\rm eff}$} &
\multicolumn{1}{c}{$\sigma(B_{\rm eff})$} &
\multicolumn{1}{c}{Sp.\ Type} \\
\hline
   358 &  3.600 & 4.140 & 2.220 & 4.29 &  2.26 &  30 & $-$369 &  82 & B8IV HgMn \\
 19400 &  3.997 & 4.148 & 2.489 & 4.10 &  2.96 & 165 &   +332 &  64 & B3V PGa \\
 26326 &  4.970 & 4.182 & 2.947 & 3.87 &  4.30 & 244 &   +119 &  75 & B5IV SPB \\
 34798 &  4.674 & 4.193 & 2.746 & 4.09 &  3.24 & 285 &   +117 &  52 & B5IV/V SPB \\
 53921 &  3.819 & 4.137 & 2.406 & 4.11 &  2.84 & 152 & $-$295 &  59 & B9IV SPB \\
 53929 &  4.201 & 4.143 & 2.641 & 3.95 &  3.61 & 256 & $-$248 & 102 & B9.5III HgMn \\
 55522 &  4.958 & 4.208 & 2.848 & 4.07 &  3.40 & 241 &   +873 &  66 & B2IV/V Bp \\
 71066 &  3.184 & 4.083 & 2.094 & 4.13 &  2.54 & 123 &    +21 &  48 & A0IV HgMn \\
 74196 &  3.504 & 4.108 & 2.275 & 4.09 &  2.79 & 148 &   +381 & 107 & B7Vp \\
 85953 &  9.182 & 4.266 & 4.130 & 3.29 & 11.41 & 870 & $-$204 &  66 & B2III SPB \\
 92287 &  6.841 & 4.215 & 3.574 & 3.51 &  7.59 & 524 &  $-$88 &  42 & B3IV SPB \\
 93030 & 13.223 & 4.477 & 4.448 & 3.97 &  6.21 & 137 & $-$189 &  78 & B0Vp \\
105382 &  4.965 & 4.211 & 2.836 & 4.10 &  3.30 & 119 & $-$430 & 102 & B6IIIe Bp \\
120709 &  4.712 & 4.214 & 2.667 & 4.25 &  2.70 &  94 &    +79 &  72 & B5III HewkPGa \\
123515 &  3.465 & 4.079 & 2.329 & 3.92 &  3.39 & 179 & $-$173 &  52 & B9IV SPB \\
131120 &  4.931 & 4.231 & 2.720 & 4.29 &  2.65 & 122 & $-$280 &  88 & B7IIIp Hewk \\
138764 &  3.582 & 4.115 & 2.308 & 4.10 &  2.80 & 111 &   +132 &  75 & B6IV SPB \\
138769 &  5.069 & 4.209 & 2.907 & 4.02 &  3.62 & 139 & $-$348 &  98 & B3IVp  \\
179761 &  4.767 & 4.114 & 2.973 & 3.55 &  6.06 & 232 & $-$267 &  63 & B8II-III \\
186122 &  3.986 & 4.112 & 2.590 & 3.85 &  3.93 & 264 &   +269 &  76 & B9III HgMn \\
209459 &  3.216 & 4.025 & 2.246 & 3.75 &  3.96 & 179 &   +144 &  60 & B9.5V \\
215573 &  3.862 & 4.144 & 2.407 & 4.15 &  2.75 & 138 &   +165 &  53 & B6IV SPB \\
221507 &  3.072 & 4.096 & 1.945 & 4.32 &  2.01 &  55 &  $-$28 &  55 & B9.5IV HgMn \\
\end{tabular}
}
\end{table}

\begin{figure}
\plottwo{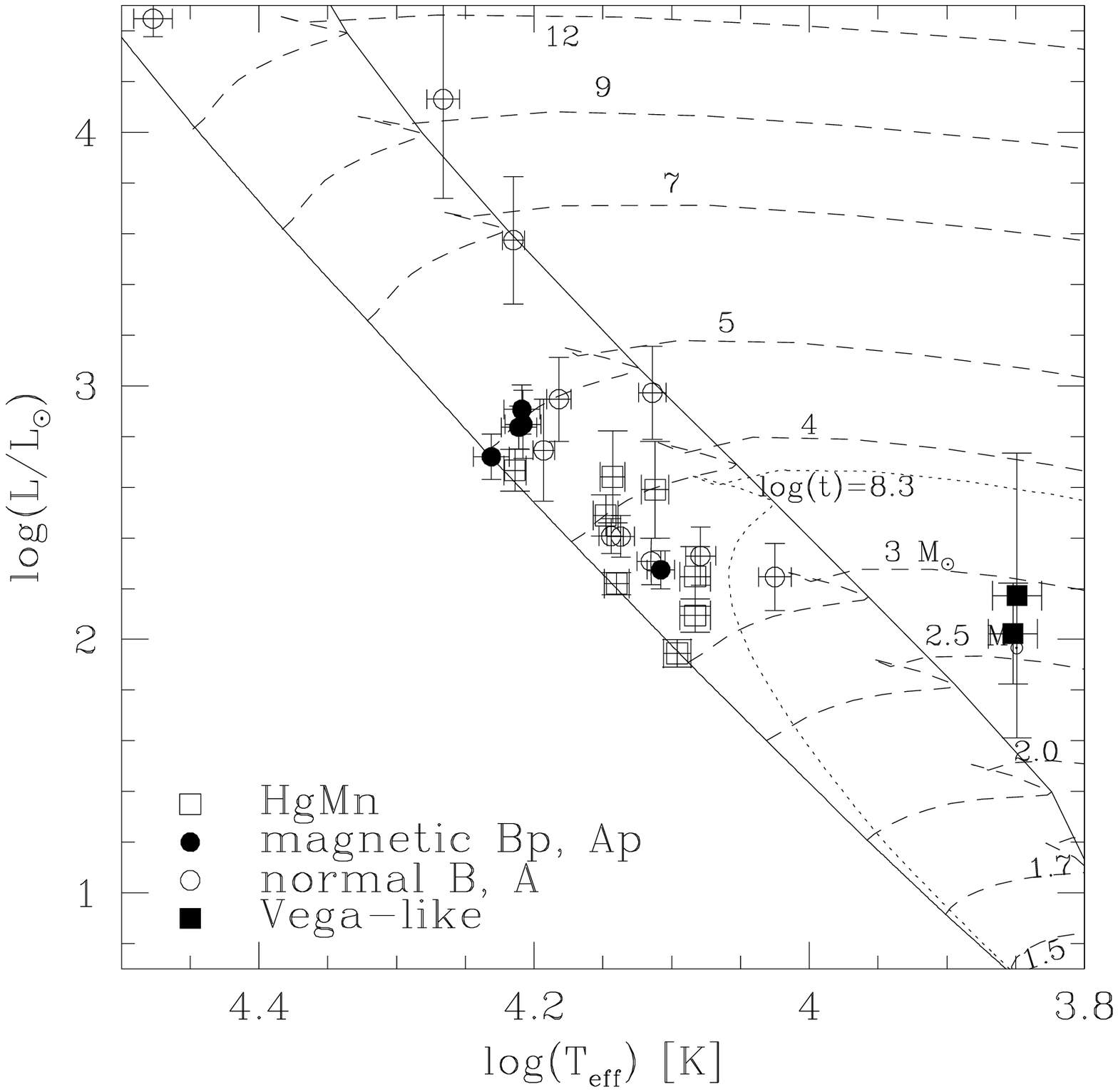}{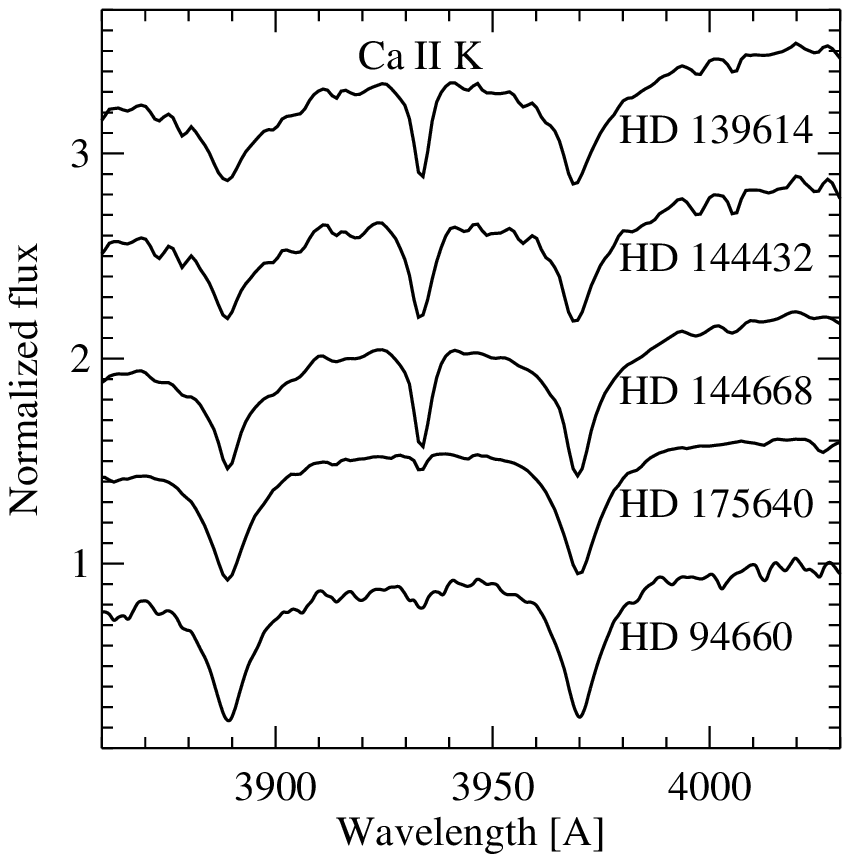}
\caption{
{\bf Left:}
The positions of the studied stars in the HR-diagram.
{\bf Right:}
Normalized Stokes I spectra of three Herbig Ae stars
(HD\,139614, HD\,144432, HD\,144668),
the HgMn star HD\,175640 and the classical Ap star HD\,94660.
The individual spectra are displaced by 0.65 with respect to each other.
}
\end {figure}

\begin{figure}
\plottwo{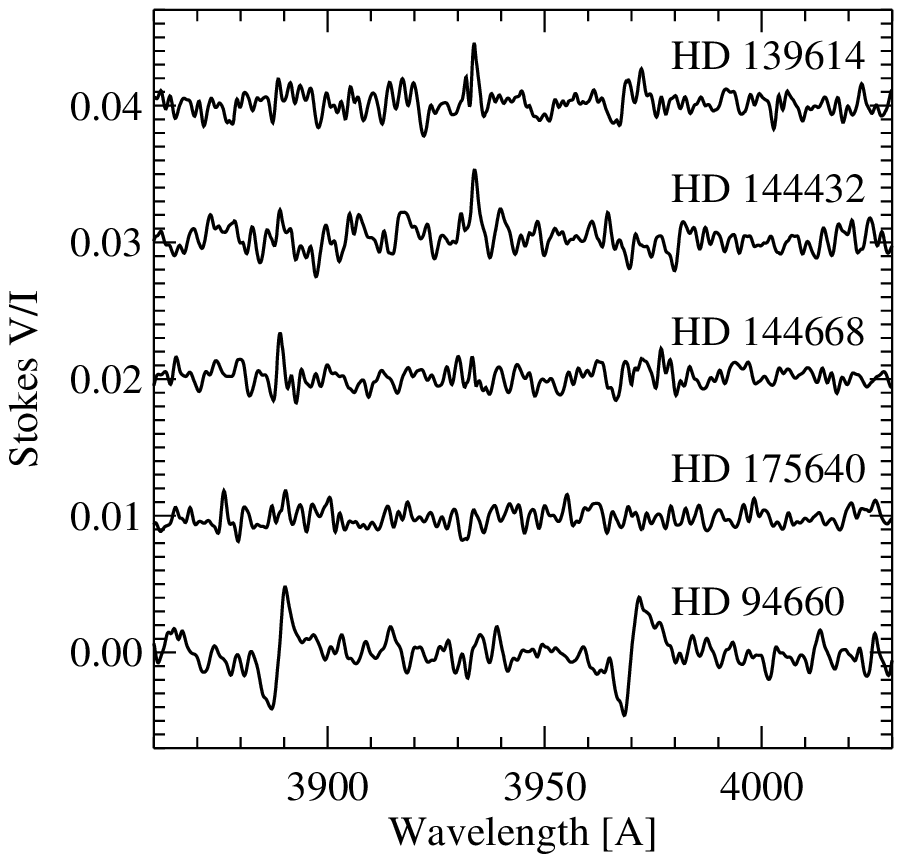}{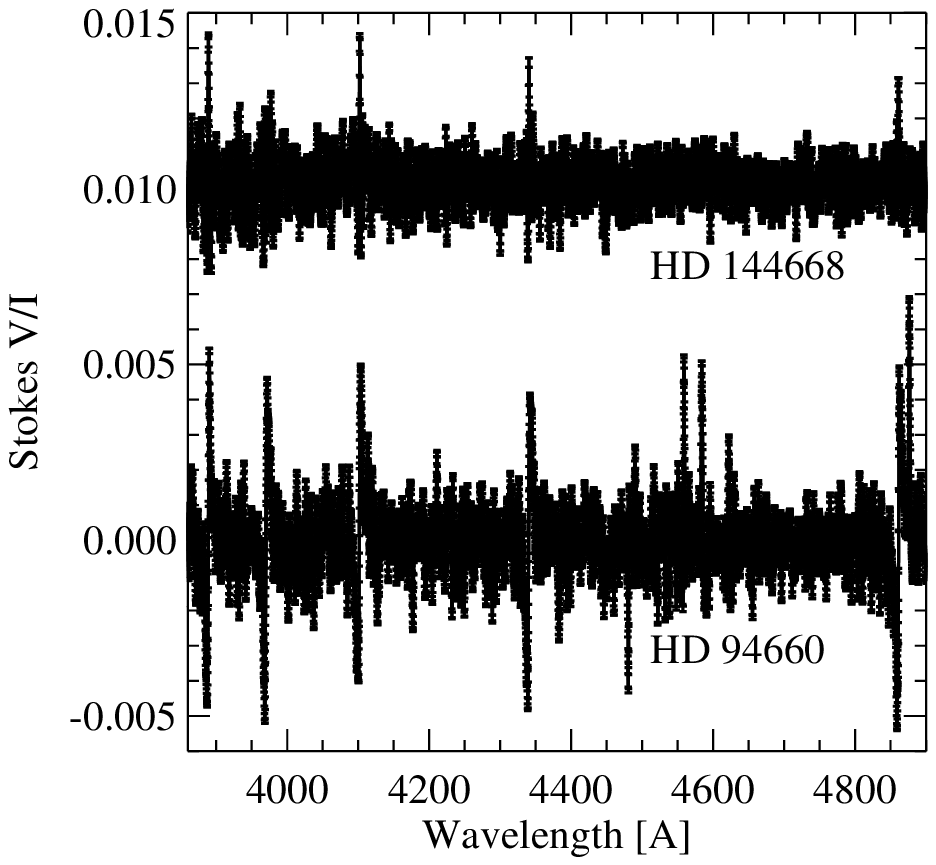}
\caption{
{\bf Left:}
Stokes V spectra of the studied Herbig Ae stars around the Ca\,II doublet.
The individual spectra are shifted by 0.01 with respect to each other.
{\bf Right:}
Stokes V spectra of the Herbig Ae star HD\,144668 and
the classical Ap star HD\,94660.
The thickness of the plotted lines corresponds to the uncertainty of the
measurement of polarization determined from photon noise.
}
\end {figure}

\section{Discussion}

Normal B, HgMn, PGa and SPB stars
are usually regarded as non-magnetic stars. The only detection of a magnetic
field in an SPB star ($\zeta$~Cas) has been presented by \citet{Ne03}.
However, the role that magnetic fields
play in the understanding of pulsational properties of SPB stars is still
unclear and further observations are needed to look for possible relations
between magnetic field and pulsation patterns.
It may be an essential clue for the understanding of the origin of the
chemical anomalies of HgMn stars that many stars of this peculiarity
type are very young and are located on the ZAMS or close to it. Previous
searches for magnetic fields in HgMn stars had shown that these stars,
unlike classical Ap stars, do not have large-scale organized fields
detectable through polarimetry. However, we were able to detect longitudinal 
fields of the order of a few hundred Gauss in two HgMn stars.
We also detected a magnetic field at 4.2\,$\sigma$ level in the
normal B-type star HD\,179761.
Three years ago we already showed for this star
evidence for a relative magnetic intensification of Fe\,II lines
produced by different magnetic desaturations induced by different
Zeeman-split components (\citet{HC01}).
As the relative intensification is roughly correlated
with the strength of the magnetic field, it is a powerful tool for detecting
magnetic fields which have a complex structure and are difficult to detect by
polarization measurements.
The intriguing discovery of mean
longitudinal magnetic fields of the order of a few hundred Gauss in a sample
of so-called "non-magnetic" stars rises a fundamental question about the
possible ubiquitous presence of a magnetic field in upper main sequence stars.
The structure of the field in these stars must be, however,
sufficiently tangled so that it does not produce a strong net observable
circular polarization signature.

In the Stokes V spectra of HD\,139614 and HD\,144432 the interesting 
fact is the presence
of circular polarization signatures in the Ca\,II~K line,
which appear unresolved at the low spectral
resolution achievable with FORS\,1 (Fig.\,2, left).
The line Ca\,II~H 
is blended with 
the Balmer line H$\epsilon$.
As the Herbig Ae stars of our sample are surrounded by circumstellar disks, models
involving accretion of matter from the disk to the
star along a global stellar magnetic field of a specific
geometry can account for the Zeeman signatures observed in the Ca\,II\,K line.
A longitudinal field at 4.8\,$\sigma$ level
has been diagnosed for HD\,139614 and the fields of the other two stars, 
HD\, 144432 and HD\,144668, have been measured at $\sim$1.6\,$\sigma$
and $\sim$2.5\,$\sigma$ levels respectively.
Given the very low  $v$\,sin\,$i$ value for
HD\,139614 with the largest magnetic field measured,
we are very likely observing this star pole-on.
For HD\,144668, viewed at edge-on, no obvious polarization signature is visible 
in the Ca\,II~K line. However, a magnetic field is  
certainly present 
since weak polarization signatures are clearly
seen in the H Balmer lines (Fig.\,2, right). 
More magnetic studies of Herbig Ae stars are needed to properly assess 
the role of magnetic  fields in the formation process of stars of intermediate 
mass.

\vfill\pagebreak
\pagebreak


\begin{thebibliography}{}
\bibitem[Hubrig, North \& Mathys(2000)]{Hu00}
Hubrig, S., North, P., Mathys, G. 2000, ApJ 539, 352
\bibitem[Hubrig \& Castelli(2003)]{HC01}
Hubrig, S., Castelli, F.\ 2001, A\&A 375, 963
\bibitem[Hubrig et al.(2004)]{Hu04b}
Hubrig, S., et al.\ 2004, Meeting on Astronomical Polarimetry, ASP, {\sl in press}
\bibitem[Neiner et al.(2003)]{Ne03}
Neiner, C., Geers, V.C., Henrichs, H.F., et al.\ 2003, A\&A 406, 1019
\end{thebibliography}
\end{document}